\documentstyle[epsf,aps,multicol]{revtex}

\begin{document}

\title{The Fractional Quantum Hall effect in an array of quantum wires}

\author{C.L. Kane, Ranjan Mukhopadhyay, and T.C. Lubensky}
\address{Department of Physics and Astronomy, University of Pennsylvania, Philadelphia, PA 19104}
\date{\today} \maketitle

\draft

\begin{abstract}
We demonstrate the emergence of the quantum Hall (QH) hierarchy in a
2D model of 
coupled quantum wires in a perpendicular magnetic field. At 
commensurate values of the magnetic field, the system can develop 
instabilities to appropriate inter-wire electron hopping processes that
drive the system into a variety of QH states.  Some of the 
QH states are not included in the Haldane-Halperin hierarchy.
In addition, we find operators allowed at any field that
lead to novel crystals of Laughlin quasiparticles.
We demonstrate that any QH state is the groundstate of a Hamiltonian that
we explicitly construct.

\end{abstract}


\begin{multicols}{2}

The rich phenomenology of the quantum Hall (QH) effect provides a fertile
setting for the study of correlated electrons\cite{QH}.
While the integer QH effect can be understood in terms of the Landau 
quantization of non interacting electrons, the fractional QH state is a 
strongly correlated quantum liquid, where electron-electron interactions 
play an essential role.  Motivated by Laughlin's original variational
wavefunction\cite{laughlin}, a number of techniques have been developed 
to describe 
the hierarchy of fractional QH states\cite{handh}, including composite fermion variational
wavefunctions\cite{jain} and Chern-Simons field theories based on 
bosons\cite{read,wenzee} or fermions\cite{hlr}.  These have lead to 
a deep understanding of the excitation spectrum of QH states
and of the structure of the QH hierarchy.

  The purpose of this paper is to develop a new formalism,
which reproduces the QH hierarchy in a model consisting of a
two-dimensional array
of quantum wires in a perpendicular magnetic field $B$. 
The model could be relevant for semiconductor
quantum wires, ropes of carbon nanotubes, and for stripes that arise in
QH systems in the higher Landau systems \cite{eisenstein1}. 
Aside from the direct relevance of the model, our calculations
provide a novel, and in many ways a simpler, approach to
describe the fractional QH effect. 
Given its success in treating the Fractional QH effect,
it is likely that our technique will 
prove useful for understanding other strongly correlated states.

We use the bosonization technique \cite{delft}, developed
for one-dimensional systems, and
relate the QH effect to coupled Luttinger liquids.
It has  been shown recently \cite{smectic,efkl,slide} that
for a range of interwire charge and current interactions, there is a
phase in which interwire Josephson, charge-
and spin-density-wave, and single-particle couplings are irrelevant.
This sliding Luttinger liquid (SLL) or smectic metal phase 
is the quantum analog
of the sliding phase found recently in DNA-lipid complexes and
stacked XY models~\cite{olt,sliding}. The SLL resembles a Luttinger
liquid, with transport properties that exhibit power-law singularities as a
function of temperature. It has also been demonstrated \cite{sondhi} that in a 
magnetic field the phase space of SLL expands considerably. 

We will show that at commensurate magnetic fields 
the SLL phase can be unstable to 
inter-wire electron hopping processes that lead the formation of an energy
gap and the QH effect.  
The presence of the $\nu=1$ QH
 state in such a model was first noted by Sondhi and
Yang\cite{sondhi}. 
Our calculation builds on this work.  We systematically
classify all electron hopping operators
and identify those that lead to QH states.  
This construction leads to QH states which are not in the 
Haldane-Halperin hierarchy\cite{handh}.
We also show that any QH state is the groundstate of a Hamiltonian that we
explicitly construct.

\begin{figure}
\par\columnwidth20.5pc
\hsize\columnwidth\global\linewidth\columnwidth
\displaywidth\columnwidth
\epsfxsize=3.0truein
\centerline{\epsfbox{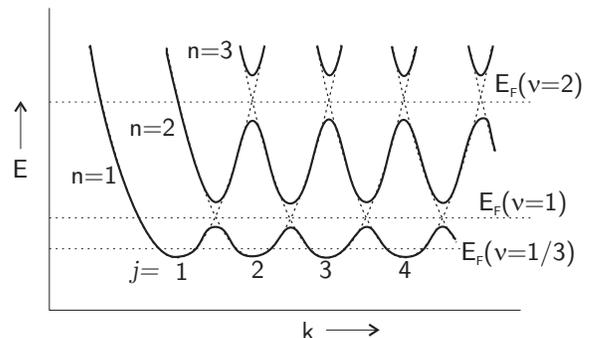}}
\caption{A schematic diagram showing how the Landau levels, denoted
by the solid curves, arise due to coupling between neighboring 
single-wire bands.}
\label{fig:Landau}
\end{figure}

We begin with a simple model of spinless electrons that ignores both
electron-electron interactions and tunneling between the wires. In
the Landau gauge $A = - B y \hat x$, the electronic dispersion has
the simple form,
\begin{equation}
 E_j(k) = {\hbar^2\over {2 m}} (k- b j)^2,
\end{equation}
where the integer $j$ labels the wires,
$b = e a B/\hbar c$, and $a$ is the separation between neighboring wires. 
This dispersion is characterized by level
crossings, where the bands associated with different wires intersect.
Tunneling between the wires couples the bands,
leading to anticrossings, as indicated in Fig.
1. The Fermi energy, $E_F$, will lie in one of these gaps when the filling
factor $\nu = 2 k_F/ b$ is an integer (here $k_F = \pi n_e$
depends on the 1D electron density $n_e$ on each wire).  In
general, the gap at $E_F$ for $\nu = N$ results from
tunneling between $N$'th neighbor wires.  At these fillings, we have
the analog of filled Landau levels and the integer QH 
effect.  It can further be observed in Fig. 1 that each filled
Landau level has a corresponding branch of gapless edge states.

Away from integer filling fractions the non-interacting theory has
gapless excitations on each wire and corresponds to a trivial
SLL fixed point.  It should
be noted that the near $E_F$ the dispersion (1) is
equivalent to what would be expected in a mean field theory of
QH stripes\cite{fandm}.  The only difference is that the left and
right moving states at $E_F$ are no longer localized on
the same wire, but rather are spatially separated.  Thus, in
addition to arrays of quantum wires what follows may also describe
instabilities of stripe phases.

With interactions, the SLL fixed point is
quite delicate, and one usually expects instabilities at low
energies.  To characterize the resulting phases we
bosonize the states near $E_F$ and write \cite{delft}
\begin{equation}
{\psi}_{R/L,j}(x) = {1 \over \sqrt{4 \pi \epsilon}} {\eta}_{R/L,j}
          e^{ i (b j \pm k_{F}) x}  e^{i \Phi_{R/L,j}(x)},
\end{equation}
where $\epsilon$ is an inter-wire cut off and $\eta$ is the Klein
factor.  The bosonic fields $\Phi_{R/L,j}$ are related to the
right and left moving electron density on each wire, $n_{R/L,j} =
\pm \partial_x \Phi_{R/L,j}/2\pi$, which satisfy the 
current algebra, $[n_{R/L,j}(x),n_{R/L,j'}(x')] = \pm i
\partial_x \delta(x-x') \delta_{jj'}/2\pi$.

The non interacting model has the
Hamiltonian density ${\cal H}_0 = \pi v_{F} \sum_j  (n_{L,j}(x)^2
+ n_{R,j}(x)^2)$, where $v_F$ is the Fermi velocity. There are 
two classes of interactions:  forward
scattering, and interchannel scattering.  The forward scattering
terms involve interactions between the densities of the various
channels, leading to a Hamiltonian which is  quadratic in the
densities,
\begin{equation}
{\cal H}_{f.s.} = {\bf n}_i(x)^T {\bf M}_{ij} {\bf
n}_j(x)
\end{equation}
where ${\bf n}_i^T = (n_{R,i},n_{L,i})$ and ${\bf M}_{ij}$ is a 2 by 2
matrix.  
${\cal H}_0 + {\cal H}_{f.s.}$ describes a  general SLL 
fixed point.  To address the the instabilities of
this fixed point, we consider interchannel scattering
interactions.  The allowed terms are built from products of 
single-electron operators, and may be classified according to the number
of times each electron operator appears:
\begin{equation}
{\cal O}_{ \{s_p^L,s_p^R \}} = \sum_{j}  \prod_p (
\psi_{R,j+p} )^{s^R_p}( \psi_{L,j+p} )^{s^L_p},
\end{equation}
where $s_{p}^{L}, s_{p}^{R}$ are integers and $\psi^s$ is taken
to mean $(\psi^\dagger)^{|s|}$ for $s<0$.  Charge conservation
requires $\sum_{p}
[s_{p}^{R} + s_{p}^{L}] = 0$, while momentum conservation requires
\begin{equation}
k_{F} \sum_{p}(s_{p}^{R} - s_{p}^{L}) + b \sum_{p} p (s_{p}^{R} +
s_{p}^{L}) = 0.
\end{equation}
Operators violating (5) will upon bosonizing using (2) have
an oscillating phase that renders it irrelevant at long
distances.

The SLL fixed point is unstable if any 
allowed interchannel scattering interactions are relevant under
the renormalization group.  This happens when the scaling
dimension $\Delta_{\{s_{p}^{R},s_{p}^{L}\}}<2$.  We
then expect the system to flow to a phase which is
characterized by ${\cal O}_{\{s_{p}^{R},s_{p}^{L}\}}$.
$\Delta_{\{s_{p}^{R},s_{p}^{L}\}}$ depends on the forward
scattering interactions ${\bf M}_{ij}$.  In principle, ${\bf
M}_{ij}$ can be parameterized with a suitable model of the
electron electron interactions.  However, ${\bf M}_{ij}$ may be
strongly renormalized by irrelevant and/or momentum non-conserving
operators, and it may not resemble the bare interactions.  Below we
argue that it is usually possible
to construct a Hamiltonian in
which a given operator $\cal O$ is the leading relevant operator.   
Our approach
is, therefore, to assume that $\cal O$ is relevant.
We then analyze the resulting strong-coupling phase.

Symmetry requires operators related by
$180^\circ$ rotation to be equivalent.  We refer to operators
which are invariant under $180^\circ$ rotation as
``non-degenerate".  Upon bosonization these interactions have the
form ${\cal H}_{\rm int} = - \sum_i u \cos \Xi_i$, where
$u$ is the magnitude of the interaction
and $\Xi_i = \sum_p (s_p^R \Phi_{i+p}^R + s_p^L \Phi_{i+p}^L) $.
For large $u$ this will tend to lock $\Xi_i$.  Since
non-degenerate operators satisfy $s_p^L = \pm s_{-p}^R$ it follows
that $[\Xi_i,\Xi_j]=0$ for all $i,j$.  Thus all
$\Xi_i$ can be simultaneously localized, and the strong coupling
phase may be described by replacing $-u \cos \Xi_i$
by $u\Xi_i^2/2$.  The Hamiltonian
is then quadratic in the boson fields and the low energy excitation
spectrum can be determined.
Lower-symmetry ``Degenerate" operators must come in
pairs.  An example is the single electron tunneling term $\sum_i
[\cos(\Phi_{R,i} - \Phi_{R,i+1}) + \cos(\Phi_{L,i} - \Phi_{R,i+1})]$,
which leads to a 2D Fermi liquid at $B=0$.
This interaction cannot be analyzed by replacing the cosine by a
square because the arguments of the cosines do not commute, so
localization is forbidden by the uncertainty principle.  
We confine our attention here to non-degenerate
operators, which can be analyzed using bosonization.

The phases described by non-degenerate operators at finite $B$ fall into two 
general categories\cite{SCops}:

(i) {\it Crystalline states}:  From (5) it can be seen that when
both $\sum_p s_p^R = \sum_p s_p^L = 0$ and 
$\sum_{p} p (s_{p}^{R} + s_{p}^{L}) = 0$, 
 the operator ${\cal
O}_{\{s_R,s_L\}}$ is allowed for any $B$.  These
operators, which independently conserve the number of right moving
and left moving electrons lead to crystalline phases of the
electrons.  A simple example is the charge density wave (CDW)
operator $\sum_i \cos[2(\theta_i - \theta_{i+1})]$, where $\theta_i =
(\Phi_{R,i} - \Phi_{L,i})/2$ is the CDW phase on each wire. When this
operator, depicted in Fig. 2b, dominates, the CDW's on neighboring
wires lock together forming a two dimensional Wigner crystal.  It
can be seen by expanding the cosine that this state has a gapless
phonon mode associated with the broken translational symmetry.
This category also includes more exotic crystals, such as the
Abrikosov flux lattice\cite{balents} (Fig. 2e) and crystals of Laughlin
quasiparticles (Fig. 2f).  These states share the common feature
that they are allowed at any $B$ and possess a gapless
phonon mode.

(ii) {\it Quantum Hall states}:  When $\sum_p s_p^R \ne \sum_p
s_p^L$ the operator ${\cal O}_{\{s_R,s_L\}}$ is allowed only at a
special magnetic field  which corresponds  to filling factor
\begin{equation}
\nu = 2 {{\sum_{p} p (s_{p}^{R} + s_{p}^{L})} \over {
\sum_{p}(s_{p}^{L} - s_{p}^{R})}}. \label {fillingfactor}
\end{equation}
The denominator of (\ref{fillingfactor})  counts the
net number of electrons backscattered from right to left moving
channels. The numerator gives the change
in the ``center of mass'' of the electrons.  By replacing
the cosine by a square, it can be established that these states 
have a gap to bulk excitations and have gapless
edge states.  The simplest example is the $\nu = 1$ QH 
state described by the operator
 $\sum_{j} \cos[\Phi_{R,j} - \Phi_{L,j+1}]$ (Fig. 2c).  This locks the 
right and left moving modes of neighboring wires, opening a gap for all
modes except for the edge mode $\Phi_{L,1}$, which remains
uncoupled. This is equivalent to the description in terms of non
interacting electrons shown in Fig. 1.

\begin{figure}
\par\columnwidth20.5pc
\hsize\columnwidth\global\linewidth\columnwidth
\displaywidth\columnwidth
\epsfxsize=3.0truein
\centerline{\epsfbox{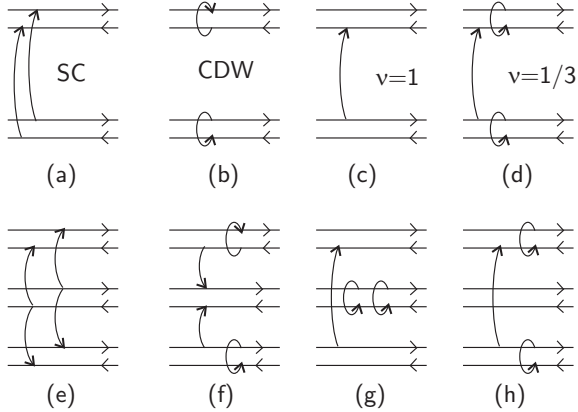}}
\caption{Pictorial representation of tunneling operators corresponding
to the following states: 
(a) Superconductor, (b) 
CDW, (c) $\nu = 1$ QH state, (d) 
$\nu=1/3$ QH state,  (e) Abrikosov flux lattice, (f) Crystal of
$\nu=1/3$ Laughlin quasiparticles, (g,h) Two distinct $\nu=2/3$ 
QH states.  } 
\label{fig:EF2}
\end{figure}

In addition to the integer QH states, which have
``single particle" energy gaps,  there are a variety
of fractional QH states whose gaps
arise from correlated tunneling processes.  The simplest 
correspond to the Laughlin sequence at $\nu = 1/m$, where $m$ is
an odd integer, and  are shown in Fig. 2d for 
$\nu = 1/3$.  An electron hops from one wire to
the next while simultaneously backscattering $(m-1)/2$ electrons
on each wire.  To establish the equivalence between the 
resulting state and the usual Laughlin states, it is necessary to
identify its topological order which may be characterized by the
set of fractionally charged quasiparticles and by 
the structure of the gapless edge states~\cite{wen}.

It is useful to transform the Hamiltonian into a form
where the edge state and quasiparticle structure are transparent.
We begin by defining $\varphi_j = (\Phi_{R,j} + \Phi_{L,j})/2$,
$\theta_j = (\Phi_{R,j} - \Phi_{L,j})/2$.  $\theta_j$ is related
to the charge density, $n_j = \partial_x\theta_j/\pi$, and
$\varphi_j$ is the conjugate phase.  
The interaction in Fig. 2d then has the form
\begin{equation} {\mathcal{O}}_{{\rm
Laughlin},m}= \sum_{j} \cos[m (\theta_{j+1} + \theta_{j}) -
\varphi_{j+1} + \varphi_{j}]. 
\label {qheoperator}
\end{equation}
We now define new right and left moving fields, $\tilde
\Phi_{R/L,j} = \varphi_j/m \pm \theta_j$ which commute with each
other and satisfy $[{\tilde \Phi}_{R/L,j}(x),{\tilde \Phi}_{R/L,j}(x')] 
= \pm i\pi {\rm sgn} (x-x')/m$.  
(\ref{qheoperator}) now becomes
$\sum_{j}\cos[m (\tilde{\Phi}_{R, j} - \tilde{\Phi}_{L,
j+1})]$.   This is similar to the $\nu=1$ integer
QH effect: left movers on wire $j+1$ lock
to right movers of wire $j$, while $\tilde \Phi_{L,1}$
decouples and remains gapless.  It is further
simplified by transforming to new charge/phase variables,
where we ``switch partners" for the right and left movers:
$\tilde\theta_{j + 1/2} = (\tilde\Phi_{R,j}-\tilde\Phi_{L,j+1})/2$,
$\tilde\varphi_{j + {1/2}} =(\tilde\Phi_{R,j}+\tilde\Phi_{L,j+1})/2$.
In terms of $\tilde\theta_{j+1/2}$,
$\tilde \varphi_{j+1/2}$ and $\tilde \Phi_{L,1}$, we have
${\cal H} = {\cal H}_{\rm b} + {\cal H}_{\rm e} +
{\tilde {\cal H}}_{\rm f.s.}$,
where
\begin{equation}
{\cal H}_{\rm b} = \sum_{j=1}^\infty v[ (\partial_x\tilde\theta_{j+{1\over 2}})^2 +
(\partial_x\tilde\varphi_{j+{1\over 2}})^2] 
- u\cos 2m\tilde\theta_{j+{1\over 2}}
\end{equation}
describes bulk states,
${\cal H}_{\rm e} = v_e (\partial_x
\tilde\Phi_{L,1})^2$ describes edge states
and ${\tilde {\cal H}}_{\rm f.s.}$ is quadratic in $\partial_x\tilde\varphi_j$,
$\partial_x\tilde\theta_j$ and $\partial_x\tilde\Phi_{L,1}$.

When $u$ flows to strong coupling $2m \tilde\theta_j$ is pinned at
a multiple of $2\pi$, and there is a gap to bulk excitations.
Solitons in which $2m \tilde\theta_j$ advances by $2\pi$
are Laughlin quasiparticles.  Since $n_j =
\partial_x\tilde\theta_j/\pi$ their charge is clearly $e/m$. To
describe the low-energy edge excitations, the gapped bulk
modes may be integrated out. The only effect of 
${\cal H}_{\rm f.s.}$ is then to renormalize the velocity $v_e$
of the edge states.  The integer $m$ which appears in the commutation
relation obeyed by $\tilde\Phi_{L,1}$ characterizes chiral 
Luttinger-liquid edge states\cite{wen} and determines the 
Luttinger liquid suppression
of the tunneling density of states, $\rho(E) \propto E^{(m-1)}$.
It should be noted that the electron creation operator for the edge states is
 $\psi_{e} \equiv  e^{i m \tilde \Phi_{L,1}} \sim \psi_{L,1}^{0}
(\psi^{0 \dagger}_{R,1} \psi_{L,1}^{0})^{(m - 1)/2}.$
The bare electron creation operator $\psi_{L,1}^0$ involves
$\tilde\varphi_{3/2}$, and is gapped.

We now address the dimension of (\ref{qheoperator})
at the SLL fixed point. 
We find a small but finite region in the space of possible interactions
in ${\cal H}_{\rm f.s.}$
where (\ref{qheoperator}) is the leading relevant operator.
However, we postpone a systematic search through parameter space
to a later publication and instead 
demonstrate that by carefully choosing a particular set of interaction
parameters we can make  (\ref{qheoperator}) as relevant as we like.
Working in the transformed representation, let 
${\tilde{\cal H}}_{\rm f.s.} = w \sum_i (\partial_x
\tilde\theta_{i+1/2})^2$.  Then the dimension of
$\cos (2m\tilde\theta_{i+1/2})$ is $\Delta= 2 m (1+w/v)^{-1/2}$.  
Thus for sufficiently
large $w$, $\Delta<2$.  Moreover, any other allowed operator
will necessarily involve
either higher powers of $\tilde\theta_{i+1/2}$ or
$\tilde\varphi_{i+1/2}$.  Since the
dimension of $\exp[ i \tilde\varphi_{i+1/2}]$ is $(1+w/v)^{1/2}/2 m$, 
these operators will be
irrelevant for sufficiently large $w$.  Thus, we have established
that it is possible to construct SLL models that
are unstable to the QH interactions and flow at low energy to the 
strong coupling QH state.  This construction
can be generalized to the more exotic QH states
considered below.

The QH states occur only at special magnetic fields, and
it is well known that disorder is crucial for the existence of
QH plateaus.  Suppose we are 
close to, but not at, a Laughlin fraction, $\nu = 1/m + \delta$.
The violation of momentum conservation in (\ref{qheoperator}) will
only become evident on a length scale $\xi \propto \delta^{-1}$.
 On shorter lengths (\ref{qheoperator}) will tend to
renormalize ${\cal H}_{\rm f.s}$, generating a term
of the form $\sum_i(\partial_x\tilde\theta_{i+1/2})^2$.  It is
thus natural to expect that for lengths larger than $\xi$ one is
in the limit of large $w$ described above.  While
$\cos[2m\tilde\theta_{i+1/2}]$ is now forbidden because
(5) is violated, a corresponding operator,
$\cos[2m(\tilde\theta_{i+1/2}-\tilde\theta_{i-1/2})]$
is allowed.  
This operator, depicted in Fig.~2f for $m=3$, describes
a crystal of Laughlin quasiparticles,
and will be most relevant for large $w$.
If this crystal is pinned
by weak disorder, then there will be a plateau in the Hall
conductance.  The pinned crystal retains the edge states of the
$\nu=1/m$ state.

    We have focussed, so far, on the $\nu=1/m$ states, which are generated
by two-wire operators.  Hierarchical QH states are generated by
operators involving more wires.   
For concreteness we consider two
three-wire operators at $\nu=2/3$: 
${\cal O}_{I}$ (Fig. 2g) and ${\cal O}_{II}$
(Fig. 2h). 
To analyze the resulting phase if either of these
 is relevant, we again write 
${\cal H} = {\cal H}_{\rm edge} + {\cal H}_{\rm blk} + {\cal H}_{\rm f.s.}$. 
As before, by adjusting ${\cal H}_{\rm f.s.}$ we can
make either ${\cal O}_{I}$ or ${\cal O}_{II}$ dominate.
The resulting phase is gapped in the bulk, 
and there are two edge modes \cite{footnote2} 
$\Phi^{({\rm edge})}_{1,2}$  which obey 
\begin{equation}
 [{\Phi}^{({\rm edge})}_{\alpha}(x), {\Phi}^{({\rm edge})}_{\beta}(x')]= 
- i\pi K_{\alpha\beta}^{-1}  {\rm sgn} (x - x').
\end{equation}
The matrix ${\bf K}$ encodes the structure of the QH states.
For ${\cal O}_{I}$, ${\bf K}$, in a diagonal basis, has $K_{11}=1, K_{22}=-3$.
This state belongs to the Haldane-Halperin hierarchy\cite{handh}
 and has two edge 
states which propagate in opposite directions.
For ${\cal O}_{II}$, ${\bf K}$ has the diagonal form $K_{11}= K_{22}=3$. 
This state, whose edge states  propagate in the same direction,
does not belong to the usual hierarchy, but rather to a more general hierarchy
considered by Wen and Zee\cite{wenzee}.  
It is related to a bilayer state in which each 
layer is in a $\nu = 1/3$ state.  These states can be distinguished by their
exponents for the temperature dependence of tunneling into the edge.

 In this paper we have developed a new formalism that allows us
to obtain the complete hierarchy of QH states with considerable
ease. However, many questions remain. It would be interesting
to use this approach to study the plateau transitions in the fractional
QH effect, which would be manifested as transitions between different
quasiparticle crystals.
In addition, we have only
considered spin-polarized electrons; it is possible that by taking into
account spin-dependent interactions we could extend our formalism
to explain, for example, the QH effect in quasi-1D conductors
\cite{FISDW}, or conductance measurements for quantum wires. 
We would also like to understand
the nature of the phases corresponding to degenerate operators,
which should include Fermi liquids of composite particles\cite{hlr}.
Finally, it would be interesting to see whether, within our formalism,
it is possible to obtain the Pfaffian state suggested for $\nu=5/2$
\cite{read2}. These form the subject for future investigations.

RM and TCL acknowledge support from the National Science
Foundation under grant number DMR00-96531.

\end{multicols}


\begin{thebibliography}{10}

\bibitem{QH} See, for example, S. Das Sarma and A. Pinzuk (ed.),
Perspectives in Quantum Hall Effects: Novel Quantum Liquids in 
Low-Dimensional Semiconductor Structures, (Wiley, New York), 1996.

\bibitem{laughlin} R.~B.~Laughlin, Phys.~Rev.~Lett. {\bf 50}, 1395
(1983).

\bibitem{handh} F.D.M. Haldane, Phys. Rev. Lett. {\bf 51}, 605 (1983);
B.I. Halperin, Phys. Rev. Lett. {\bf 52}, 1583 (1984).

\bibitem{jain} J. K. Jain, Phys. Rev. Lett. {\bf 63}, 199 (1989).

\bibitem{read} N. Read, Phys. Rev. Lett. {\bf 65}, 1502 (1990).

\bibitem{wenzee} X.G. Wen and A. Zee, Phys. Rev. B {\bf 46}, 2290 (1992).

\bibitem{hlr} B.I. Halperin, P.A. Lee and N. Read, 
Phys. Rev. B {\bf 47}, 7312 (1993).

\bibitem{eisenstein1} K. B. Cooper, M. P. Lilly, J. P. Eisenstein, 
L. N. Pfeiffer, and K. W. West, Phys. Rev. B {\bf 60}, R11285 (1999). 


\bibitem{delft} See J. von Delft and H.~Schoeller, 
Annalen der Physik {\bf 7}, 225, (1998). 



\bibitem{smectic} S.A. Kivelson, V.J. Emery, and E. Fradkin,
Nature (London) {\bf 393}, 550 (1998); E. Fradkin and S.A. Kivelson,
Phys. Rev. B {\bf 59}, 8065 (1999).


\bibitem{efkl} V.J.~Emery, E.~Fradkin, S.A.~Kivelson, and T.C.~Lubensky,
Phys.~Rev.~Lett. {\bf 85}, 2160 (2000); A.~Vishwanath and D.~Carpentier, 
Phys.~Rev.~Lett. {\bf 86}, 676 (2001); R.~Mukhopadhyay, C.L.~Kane, and 
T.C.~Lubensky, Phys.~Rev.~B {\bf 63}, 081103(R) (2001).


\bibitem{slide} R.~Mukhopadhyay, C.~L.~Kane, and T.~C.~Lubensky,
Phys. Rev. B {\bf 64}, 045120 (2001).


\bibitem{olt}
 C.S.O'Hern, T.C. Lubensky and J.Toner, Phys. Rev. Lett.
{\bf 83}, 2745 (1999).


\bibitem{sliding}
L. Golubovic and M. Golubovic, Phys. Rev. Lett. {\bf 80},
4341 (1998). Erratum {\it {ibid.}} {\bf 81}, 5704 (1998).
C.S. O'Hern and T.C. Lubensky, {\it {ibid}} {\bf 80}, 4345 (1998).


\bibitem{sondhi} S.L.~Sondhi and K.~Yang, Phys.~Rev.~B {\bf 63}, 
054430 (2001).

\bibitem{fandm} A.H. MacDonald and M.P.A. Fisher, 
Phys. Rev. B {\bf 61}, 5724 (2000).

\bibitem{SCops} In addition there is a class of superconducting operators 
allowed at $B=0$ (Fig. 2a) as well as operators that are not 
allowed for any finite $B$.

\bibitem{balents} L. Balents and L. Radzihovsky, 
Phys. Rev. Lett. {\bf 76}, 3416 (1996).


\bibitem{wen} X.G. Wen, Phys Rev. B {\bf 43}, 11025 (1991); Phys. Rev. Lett.
{\bf 64}, 2206 (1990).



\bibitem{footnote2} In general, for hopping operators involving upto
$n$ neighboring wires, the corresponding QH state will have
$n-1$ edge states.    


\bibitem{FISDW} V.~M.~Yakovenko and H.~S.~Goan, 
Journal de Physique I (France) {\bf 6}, 1917 (1996).


\bibitem{read2} N.~Read, Physica~B {\bf 298}, 121 (2001).


\end{thebibliography}
\end{document}